# Assessing the Effectiveness of Using Live Interactions and Feedback to Increase Engagement in Online Learning

Beth Porter[1], Burcin Bozkaya[2]

**Abstract**

In-person instruction for professional development or other types of workplace training provides a social environment and immediate feedback mechanisms that typically ensure all participants are successful. Online, self-paced instruction lacks these mechanisms and relies on the motivation and persistence of each individual learner, often resulting in low completion rates. In this study, we studied the effect of introducing enabling tools and live feedback into an online learning experience on learner performance in the course, persistence in the course, and election to complete supplemental readings and assignments. The findings from our experiments show positive correlations with strong statistical significance between live interactions and all performance measures studied.

Research funded by the National Science Foundation, award number #1843391.

## 1. Introduction

Achieving and maintaining consistent student engagement can be difficult in online learning contexts (Kizilcec and Halawa, 2015; Allen and Seaman, 2015; Greene et al., 2015). There are many factors that impact learner engagement: course structure, amount of lecture time, user experience of the learning management system, and countless other variables. However, online educators and scholars of online learning have consistently found evidence that peer interaction and learner-instructor interaction are strongly related to learner engagement (Bryson, 2016; Garrison and Cleveland-Innes, 2005; Greene et al., 2015; Richardson and Swan, 2003; Yamada, 2009). Several studies have found that measures of social presence, a theoretical construct originating from communications studies, are significantly related to learners' overall outcomes and perceived experience (Richardson and Swan, 2003). Social presence refers to how "salient" an interpersonal interaction is — how readily one feels like they are "present" with another.

To achieve high social presence between students, researchers have discovered that the modality of online communication is extremely important (Yamada, 2009; Kuyath, 2008; Genevieve and Johnson, 2006). The modality of communication refers to the types of channels that are used: text chat, forums, voice chat, video conference, etc. In communications, human-computer interaction, and online education literatures, it has been shown that richer modes of communication between conversing members lead to higher rates of trust, cooperation, engagement, and social presence (Bos et al., 2002). For example, video conference, which includes real time video and voice, is a richer mode of communication than asynchronous forums, and has been shown to lead to higher rates of cooperation than text chat (Bos et al., 2002).

---

[1] Riff Analytics, 44 Fenno Rd, Newton, MA, beth@riffanalytics.ai
[2] New College of Florida, 5800 Bay Shore Rd., Sarasota, FL 34243. bbozkaya@ncf.edu

Part of the innovation at the core of our study is the burgeoning science of quantitatively analyzing human social interaction. This small but exciting subfield of computational social science has shown immense promise in developing machine understanding of human interaction (Woolley et al., 2010). By instrumenting many different types of communication, such as face-to-face physical conversation, video conference, and text chat, researchers have been able to predict group performance on a variety of tasks.

These advances along with other successful projects provide strong evidence that the interaction patterns of people carry rich signals that correlate with a variety of outcomes important to personal growth, learning, and team success. By using modern machine learning methods and statistics of conversation patterns — who spoke when, for how long, etc — even without knowing the content of these conversations, researchers have been able to predict speed dating outcomes, group brainstorming success, and group achievement on a test of "collective intelligence" (Pentland et al., 2006, Woolley et al., 2010; Dong and Pentland, 2010; Jayagopi et al., 2010; Kim et al., 2008; Dong et al., 2012a; Dong et al., 2012b). By developing more sophisticated computational models of human interaction, it is possible to infer social roles, such as group leaders or followers (Dong et al., 2013).

These signals are not isolated to face-to-face communication. Analysis of text-based modalities can also provide strong insight. Researchers have for years investigated the structure of organizations using historical data of company emails. Our recent work on analyzing the communication patterns of students in an online class suggests that text chat communication patterns are at least as strong a signal as student grade when predicting if students will lead group projects or attain high achievement (Sun et al., 2018).

To show the effectiveness of online learning systems that contribute to strong learning outcomes and learner performance via tools that foster participation and interaction, we present an experimental study in this paper conducted with an online class. Our learning platform, which is used to conduct our experiments, allows students to interact with one another using a variety of modalities including video chat and text chat. In addition, students get continuous feedback on their participation and engagement in the learning experience via a tool we named Meeting Mediator (MM). MM essentially provides feedback on how the class participant is doing in terms of various measures of engagement such as speaking time, turn-taking, influencing or affirming other participants, interrupting other speakers, etc.

In our experiments, we seek to find correlations between learner performance measured with a number of metrics, and the learner's usage of our enabling tools and live feedback. The metrics we use include learner performance in the course (pass/fail or certificate earned), persistence in the course, and election to complete optional readings and assignments. Our experiments reveal strong positive correlations (as high as Pearson's r = 0.54) with strong statistical significance between all of these output measures and the usage of our learning platform with the MM. These results indicate a strong first step towards implementing intelligent systems that "sense" participant engagement and motivation from various signals in a live learning session, which, in the future, are likely to include various components of real-time signal processing and participant nudging.

In what follows, we present our methodology including our experimental approach, followed by the presentation and discussion of results obtained.

## 2. Materials and Methods

In order to have a data collection and analytics infrastructure for carrying out experiments on the effects of real-time interactions (video and text chat experiences) and associated interventions (human and machine) on peer learning in online learning environments, we first describe our user-facing applications, supporting backend tools and the metrics we designed and employed for performance assessment. Then, we present our experiment design.

### 2.1 Riff Platform and its User-Facing Applications

Our team developed four foundational user-facing applications and supporting backend tools as part of the Riff Platform. These are:

- Riff Video Chat (with Meeting Mediator)
- Riff Metrics dashboard
- Riff Text Chat (which includes Riff Video and Riff Metrics)
- Riff Diagnostic (which includes Riff Video)

These applications and tools were built for both commercial and research purposes, cloud-deployed in largely standardized ways, but customized over time to meet the changing needs of Riff customers and modified research goals. (An open version of Riff Video, called Riff Remote, is available here: [https://my.riffremote.com/](https://my.riffremote.com/).

#### 2.1.1 Riff Video Chat

Riff Video Chat (Figure 1) is an in-browser video chat application that uses the computer's microphone to collect voice data for analysis. Riff Video Chat is present in all manifestations of the Riff Platform.

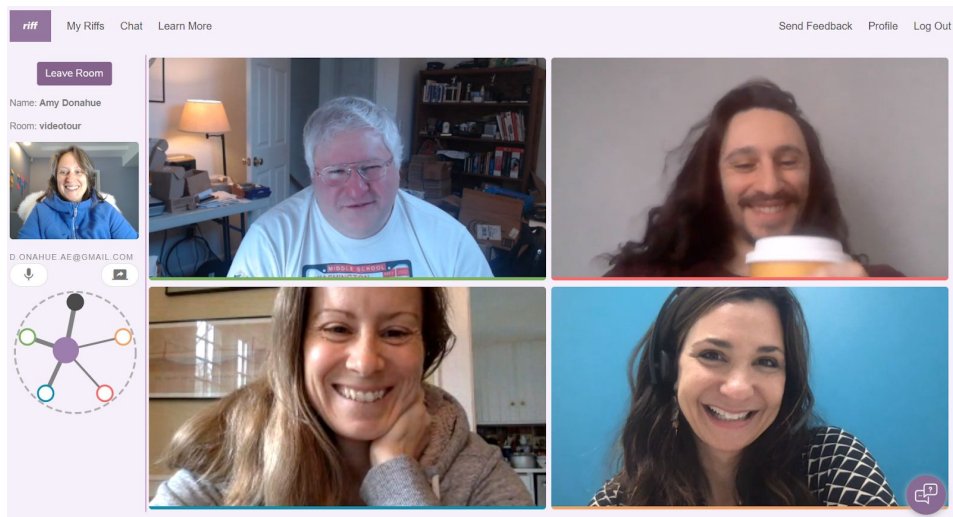

*Figure 1. Riff Video browser-based video conferencing client*

Riff Video does not require plug-ins or a local application in order to run. Participants navigate to a URL (either directly by entering it in the address bar or through authenticated redirect from another application), enable their camera and microphone, and then enter the video chat room. Other features of the video chat include screen sharing, microphone muting, and the ability to load a document (typically one with shared access) side-by-side with the video on screen.

A key element of Riff Video is the Meeting Mediator (MM) (Figure 2), which gives participants of the video chat real-time feedback about their speaking time. Specifically, it provides three metrics:

- **Engagement**, as indicated by the shade of purple of the node in the middle of the visualization, which shows the total number of turns taken in running 5-minute intervals throughout the chat — dark purple means lots of turns; light purple means fewer turns
- **Influence**, as indicated by the location of the purple node, which moves toward the person who has taken the largest number of turns in running 5-minute intervals throughout the chat
- **Dominance**, as indicated by the thickness of the grey "spokes" running from the central purple node out to each of the participant nodes, which indicates the number of turns people have taken in running 5-minute intervals

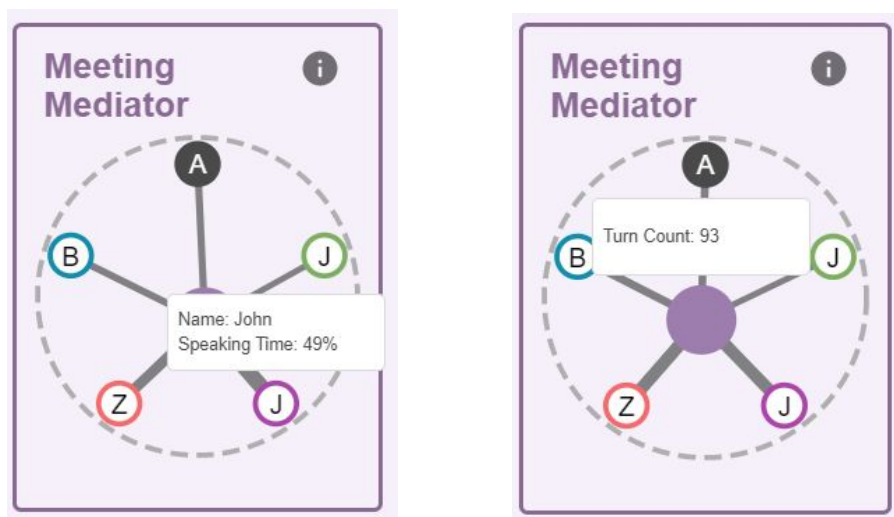

Figure 2. Hover states of the Meeting Mediator showing different metrics

These metrics are continuously updated throughout the video chat, lagging one or two seconds behind the live conversation.

2.1.2 Riff Metrics

Immediately after a Riff Video Chat meeting, Riff Metrics (Figure 3) appear on screen to show participants measures associated with the meeting, specifically:

- **Speaking Time** — a chart showing the percentages of turns taken by each participant
- **Pairwise Comparisons** — each of the following charts shows the pairwise comparisons of the individual participant with respect to each other participant, to show the balance of each pair in their interactions:

- ○ **Influences** — a chart showing who spoke after who in the video chat, which aggregates data across all types of spoken events
- ○ **Interruptions** — a chart showing interruptions (when someone cuts off another person's speaking turn)
- ○ **Affirmations** — a chart showing affirmations (when someone has a verbalization, but does not cut off another person's speaking turn)
- **Timeline** — a view of the video chat meeting showing speaking times for all participants, and is decorated with "your" and "their" interruptions, affirmations, and influences

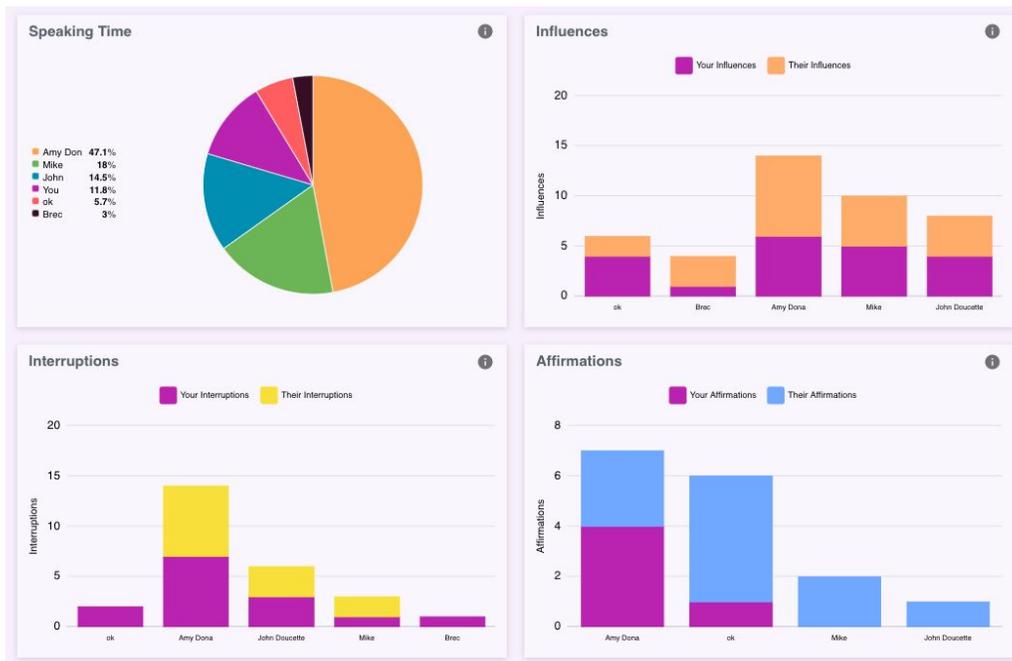

*Figure 3. Meeting metrics (speaking time, influences, interruptions and affirmations)*

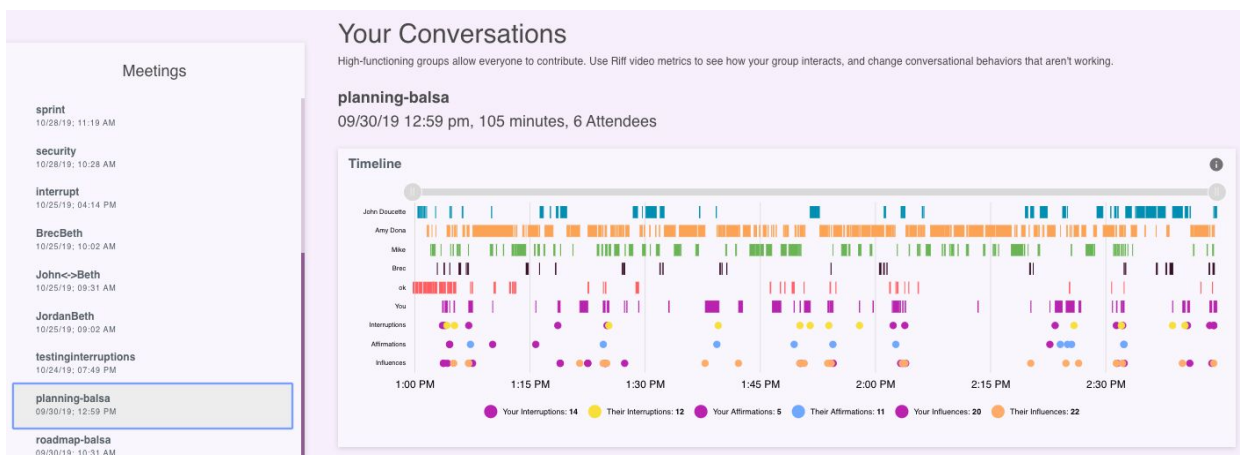

*Figure 4. Meetings over time (left) and meeting timeline (right)*

The Riff Metrics view also shows meeting history — metrics for all video chats over time — (Figure 4) and can be accessed independently of the video chat itself.

2.1.3 Riff Text Chat

Riff Text Chat (also known as Riff EDU) is an environment that allows people to text chat with one another in open (usually topical) channels, private channels, and direct messages (Figure 5).

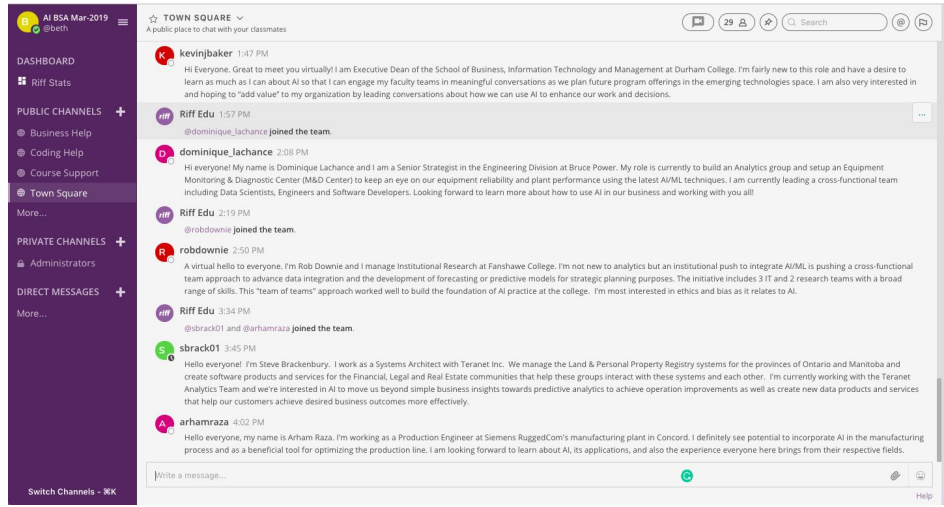

*Figure 5. Riff Chat channel-based text chat environment*

Riff Text Chat is built on top of the Mattermost platform, which is an open source project (https://mattermost.com/). The Riff team developed a customized version of this platform which has three key additions (and several smaller modifications):

- **LTI integration**, which allows the learner to move seamlessly from their learning environment (Canvas, Open edX, etc) using the learning tools interoperability (LTI) protocol for authentication
- **Riff Video integration**, which allows the learner to start an ad hoc Riff Video chat session from the Riff Chat environment
- **Riff Metrics integration**, which shows the learner views of their Riff Video meeting metrics within the Riff Chat environment, as well as additional text chat data and recommendations

2.1.4 Riff Diagnostic

To enable specific research activities, namely a controlled trial to test the effect of the Meeting Mediator on team performance, the Riff team combined the Riff Video Chat experience with surveys and complex group decision-making tasks. Specifically, Riff Diagnostic provided a self-contained experience that allowed the Riff team to collect 4 different types of measures:

- Socio-emotional acuity of each individual participant, using a survey with a modified version of the "Reading the Mind in the Eyes" test (http://socialintelligence.labinthewild.org/mite/)
- Patterns of group interaction, using Riff Video metrics (with and without the Meeting Mediator in an AB, BA, AA, and BB pattern over two tasks)
- Performance as a group, using two (2) rubric-based moral reasoning tasks
- Social presence, using a short, post-video chat survey

## 2.2 Experiment Design

To show the effectiveness of our video conferencing environment and our nudging tool, the Meeting Mediator (MM), we conducted an experiment with a major client of Riff Analytics in Canada. In this experiment, test subjects were recruited to participate in an online course called "AI Strategy and Application" from a government subsidized incubator of AI and other tech companies headquartered in Toronto, Ontario, Canada. Students elected to take the course as a professional development experience, particularly driven by interest in learning how to start an AI initiative either as an intrapreneurial or entrepreneurial activity. In most cases, the learners were paying for the course, but in some cases, either the course had been paid for by their employer or they were given free access to the course. Learners were either students in advanced degree programs or full-time professionals taking the course as a supplemental learning experience.

The 8-week course was delivered on the Open edX platform with Riff serving as the communications and collaboration platform for instructor and learner interactions, both structured and ad-hoc. Learners had opportunities to view videos and read original materials on how AI is applied to business problems, tackle hands-on coding exercises to learn basic machine learning techniques in a sandbox environment, and take assessments at the end of each learning unit.

The participants were divided into groups throughout the course to facilitate the following two types of activities:

1. **Brainstorming activities** (in pre-set groups of 4-6 people) — Each week during the first four weeks of the course, people met on Riff Video in their Peer Learning Groups (PLG) to collaborate on the assignment, and then create a shared submission.
2. **Capstone project collaboration** (in self-selected groups of 4-6 people, plus a mentor) — Each week during the last four weeks of the course, people met on Riff Video in their Capstone Groups to collaborate on developing a venture plan based on the winning pitches submitted by their peers in the course.

Learners were left to schedule video meetings with their group members on their own, and the course was time-released (one week of material at a time), but self-paced. However, the course support staff did participate in the first meeting of each group, just to make sure that people didn't have technical issues and were able to make the video chats happen without issue.

The main goal of our experiment was to analyze the relationship between using Riff Video and Riff Chat, and various outcomes among learners in the course. Furthermore, we aimed to analyze differences, if any, between early users (those who completed at least the first half of the course) versus the rest, to understand the effects of early usage on retention and grades.

Note that in this experiment, the Meeting Mediator (MM) was always present for Riff Video chats. Our rationale for that choice was that if our hypotheses that Riff Video with MM is positively correlated with any of performance, persistence, or satisfaction in online courses, then having it present for some students and not for others in a paid, graded learning experience *might* disadvantage the control population and be deemed inequitable.

In our experiment, we set out to test a number of hypotheses and hence show the (positive) relationship (or its lack thereof) between the use of Riff Video and various outcomes. Specifically, we explored the following questions:

1. Did students who used Riff Video Chat more often receive higher grades in the course?
2. Did students who used Riff Video Chat more often complete the course at a higher rate?
3. Is there a strong positive association between participating in additional Riff Video Chat during the first 4 weeks of the course and earning a certificate?
4. Is there a strong positive association between participating in additional Riff Video Chat during the first 4 weeks of the course and receiving higher grades at the end of the course?
5. Is there a strong positive association between participating in additional Riff Video Chat during the first 4 weeks of the course and completing more of the optional course assignments?
6. Is there a strong positive association between participating in additional Riff Video Chat during the first 4 weeks of the course and pitching a capstone project topic?

The first two research questions are explored for students who completed the entire course (n=62); we removed the Riff chat records and performance output results for students who dropped part way. The remaining questions are explored for the entire cohort students of n=83. These students all started the course, but some of them dropped out and, hence, did not receive a final grade for completion or a certificate.

In our analysis of our experiment results, we used the following constructs for reporting and interpreting our figures:

**Correlations:** we calculate the degree to which people who do one thing more (i.e. use Riff Video chat), also tend to do something else more (e.g. receive higher grades). Correlations range between -1 and +1. A correlation of 0 (or near-zero) means there is no discernable relationship between the two outcomes that are measured. We note, however, that the correlations reported in this study cannot be directly used to imply *causality*; they are merely indicators that two phenomena occur together (in a positive or negative relationship) or not.

**Odds Ratios:** we calculate a change in the ratio of the odds of an event occuring when some change is made to a person. They range from 0 to infinitely large. A ratio above 1 means
1 that the odds improve when the change is made. A ratio below 1 means the odds get worse. A ratio of 1 means the odds didn't change at all. As an example, if the odds of passing the course for a student who never used Riff were 1 to 5, and the odds of passing among students who used Riff exactly once were 3 to 7, then the odds ratio associated with exposing a student to one additional Riff call is [(3/7):(1/5)] = 15/7, which is a little over 2. This would mean that the odds of the student passing the course roughly double when they are exposed to one additional Riff call. Notice that this is not the same as saying they are twice as likely to pass the course. It says the odds got twice as good.

**Significance**: this captures the idea that, just because we see a strong relationship (i.e. a high correlation) between some behaviors in a past course, it might not be safe to claim that future participants will see it too. This might be because the relationship is strong but "noisy"; because it is only present among a few students; or because one has looked for so many relationships that one is bound to find something. The 'p-value' is used to quantify the significance of a relationship. A separate "Significance" column is placed next to any p-value we report.

Finally, we conducted an exit survey for all students who completed the course and a separate survey for those who failed to complete it. While the results of this survey are not directly related to the figures we report in the next section, we report the list of questions and a few key results we collected from these surveys in the Appendix.

The following section provides the results of our experimental study to answer the questions above, including correlation values, odds ratios and significance values.

## 3. Results

In this section, we report the results of our experimental study for students who did participate in the course for its entire duration (questions 1-2, n=62) and for all students who started with the course, even if they dropped out (questions 3-6, n=83) separately.

### 3.1 Results for Questions 1-2 for students who completed the course

To answer the first question in our research study ("Did students who used Riff Video Chat more often receive higher grades in the course?"), we considered the following output variables:

- Final grade earned
- Coding exercise grade
- Capstone exercise grade
- Collaboration exercise grade
- Pitch video completion

For the second question ("Did students who used Riff Video Chat more often complete the course at a higher rate?"), we simply considered whether or not the student earned his/her certificate of completion in the course. The following table shows the correlation between each output variable listed above and the input variable of "Riff Calls made", which is the number of times a student has connected to the Riff Video Chat. The correlation values were further corrected using Holm's method to maintain a FWER of 0.05.

Table 1. Correlation between Riff Video usage and various performance variables.
*$p < 0.5$, **$p < 0.01$, ***$p < 0.001$

| Attribute | Correlation to # Riff Calls Made | n | p | Significance |
|---|---|---|---|---|
| Final Grades | 0.50 | 62 | 1.56e-04 | *** |
| Coding Exercise Grades | 0.41 | 62 | 2.54e-03 | ** |
| Capstone Exercise Grades | 0.49 | 62 | 1.56e-04 | *** |
| Collaboration Exercise Grades | 0.27 | 62 | 2.94e-02 | * |
| Pitch Video Completion | 0.37 | 62 | 5.25e-03 | ** |
| Certificate Earned | 0.50 | 62 | 1.56e-04 | *** |

We further calculated the odds ratio for two of these variables, as shown in Table 2. Here, the "Grades" variable was created as a binary variable indicating the student has received (or not) a passing grade based on his/her final grade.

Table 2. Odds ratios for the effect of attending one additional Riff video call on final grade and certificate attainment obtained by fitting a logistic regression model.
*p < 0.5, **p < 0.01, ***p < 0.001

| Attribute | Odds Ratio | n | p | Significance |
|---|---|---|---|---|
| Grades | 1.23 | 62 | 6.92e-03 | ** |
| Certificate Earned | 1.35 | 62 | 3.96e-04 | *** |

The relationship between these two variables and our input variable ("Riff calls made") is further illustrated in Figures 6 and 7.

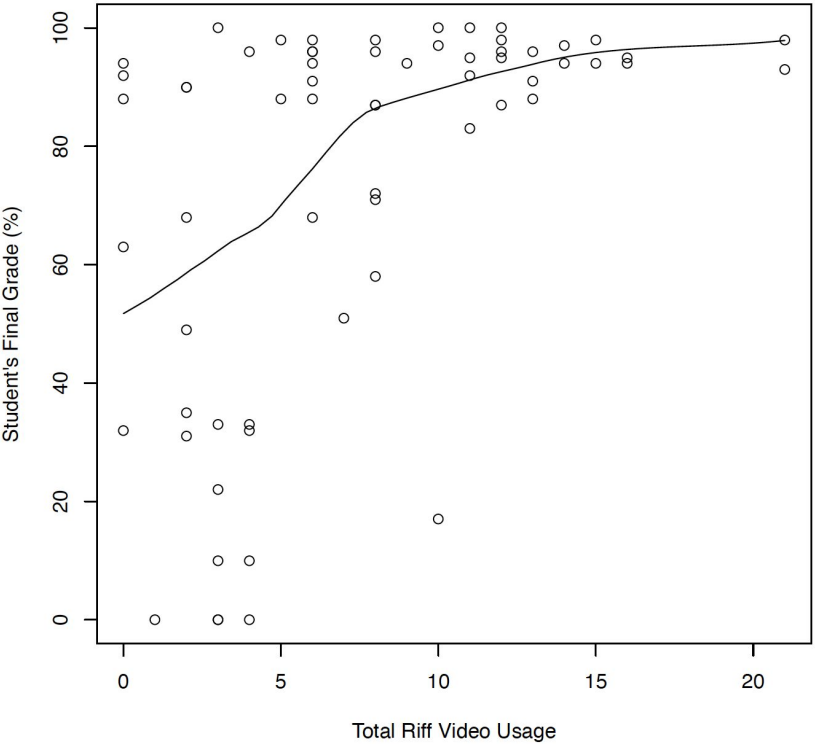

*Figure 6. Relationship between Riff Video usage and students' final grades.*

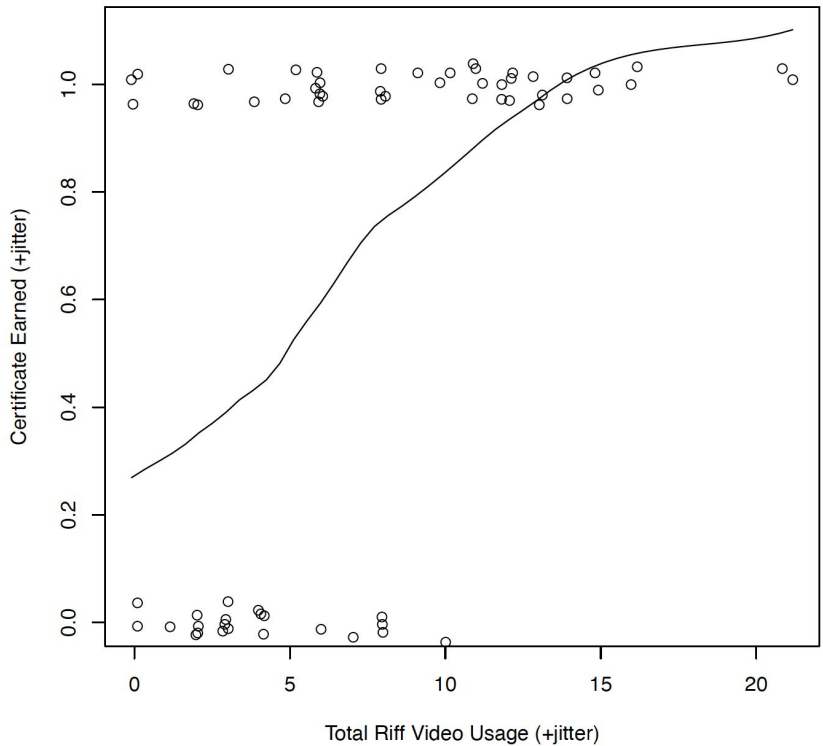

*Figure 7. Relationship between Riff Video usage and certificate achievement*

## 3.2 Results for Questions 3-6 for all students

We explored the remaining research questions regarding the early usage (first 4 weeks) of Riff Video using data we collected from all students in this course. We note again that some students completed the course only partially and dropped out some time during the course period. They did, however, complete the first 4 weeks of the course. Here we used the start of the Capstone Project as the cutoff date for tracking how often the students were present in channels in which Riff Video chats were offered. We correlated this variable to the same performance variables we considered in Section 3.1 to explore the research questions 3-6:

Tables 3 and 4 show the correlation values between the variables similar to the way they are presented in Section 3.1. All correlations were computed with Pearson's method, and *p* values were corrected again using Holm's method to maintain a FWER of 0.05.

Table 3. Correlation between early Riff Video usage and various performance variables.
*p < 0.5, **p < 0.01, ***p < 0.001, ****p<0.0001

| Attribute | Correlation to # Riff Calls Made | n | p | Significance |
|---|---|---|---|---|
| Final Grades | 0.54 | 83 | 5.40e-07 | **** |
| Coding Exercise Grades | 0.42 | 83 | 5.22e-05 | **** |
| Capstone Exercise Grades | 0.50 | 83 | 3.89e-06 | **** |
| Collaboration Exercise Grades | 0.52 | 83 | 2.07e-06 | **** |
| Pitch Video Completion | 0.45 | 83 | 3.95e-05 | **** |
| Certificate Earned | 0.50 | 83 | 3.89e-06 | **** |

Table 4. Odds ratios for the effect of attending one additional Riff video call on final grade and certificate attainment obtained by fitting a logistic regression model.
*p < 0.5, **p < 0.01, ***p < 0.001, ****p<0.0001

| Attribute | Odds Ratio | n | p | Significance |
|---|---|---|---|---|
| Grades | 1.79 | 83 | 1.03e-04 | *** |
| Certificate Earned | 2.00 | 83 | 1.96e-05 | **** |

The relationship between these two variables and our input variable ("Early Riff Video usage") is further illustrated in Figures 8 and 9. Red points in these figures indicate students who dropped the course. Note that in one case, a student indicated they were dropping the course, but then completed all work and earned a certificate.

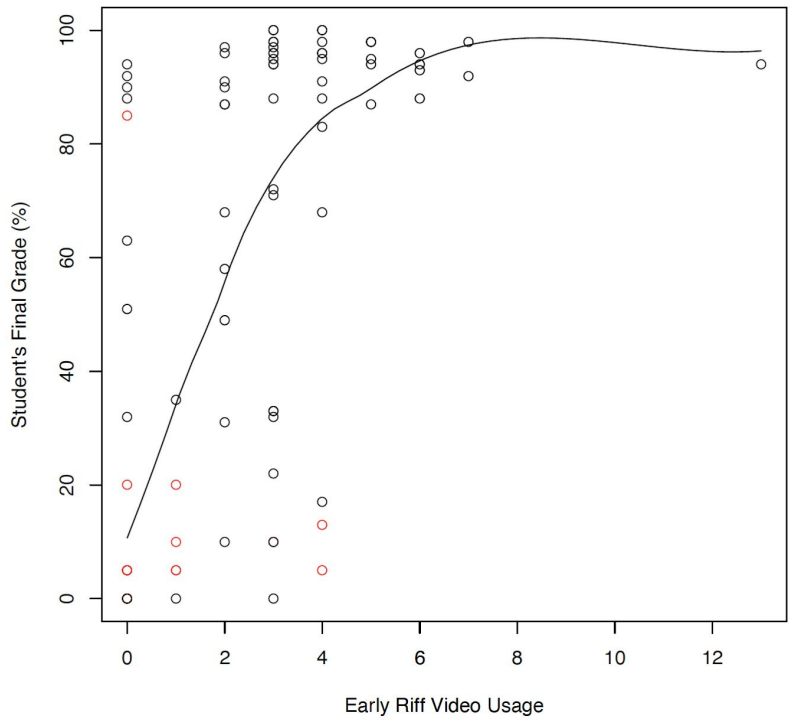

*Figure 8. Relationship between early Riff Video usage and students' final grades.*

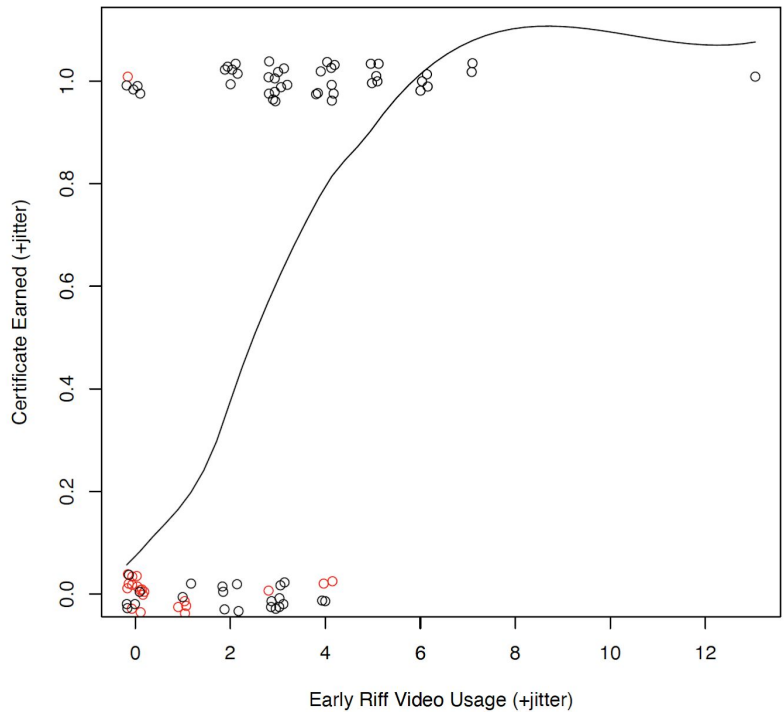

*Figure 9. Relationship between early Riff Video usage and certificate achievement.*

## 4. Summary and Discussion

### 4.1. Summary

The results of the experiment may be summarized as follows:

- Each additional Riff Video Chat made during the first 4 weeks of the course predicts a doubling of the odds that the student will receive a certificate of course completion.
- Each additional Riff Video Chat made during the first 4 weeks of the course predicts an increase in the odds of receiving a high grade by 79%.
- Most benefits are accrued after participating in the first 4-5 Riff Video chats. Students who participated in more than 4 calls (an average of 1 per week) received final grades 80% higher than those who did not, and were twice as likely to earn a certificate.

### 4.2. Discussion

In our analysis of experiment results, for the first group that completed the entire course, we found strong correlations between the time (in minutes) spent using Riff Video to communicate with peers, and every outcome of interest as shown in Table 1. Particularly strong correlations were found between students' final grades and attainment of course certificates, which are measurements of strong commercial significance. Because this is an observational design, these results do not establish a causal relationship between usage of the Riff Platform and these outcomes, but they strongly suggest that such a relationship may exist. We found that the relationship between Riff Video usage and both grades and course completion was well fit by a logistic curve, as shown in Table 2 and Figure 7.

The reported odds ratios as extracted (and reported in Table 2) from the coefficients of the logistic regression model fit suggest results that are statistically significant. The reported values correspond to increases in the odds of a student receiving higher grades by a factor of 23% for each additional hour spent using Riff Video and in the odds of passing the course by 35%. The analysis suggests that almost all of the improvements in outcomes are realized with a total exposure of approximately 15 hours over the course.

With the second group of students who remained enrolled during at least the first quarter of the course, the results summarized in Tables 3 and 4 demonstrate that early usage of Riff platform has an even stronger relationship to the outcomes of interest than usage over the course as a whole. Particularly notable is the effect size for each additional hour of exposure to Riff Video during the first half of the course, which is associated with a doubling of the odds that the student completes the course and earns a certificate. Although this is an observational experiment, it again strongly suggests that the Riff platform may be having a significant impact on students' course completion rates and participation.

These experiments and the corresponding results, though they signal strong correlations between the usage of Riff Platform and all measured output variables, are not without limitations. Clearly, the results are based only on this set of experiments with this cohort of students in a single course. It is better to repeat these experiments in a variety of settings (different cohorts in different countries in different types of learning environments and tasks) to see if the results can be generalized. Secondly, we did not have any control variables (demographic variables such as age, gender, prior job experience or other variables like prior online learning experience or prior performance in similar online or offline courses) on the subjects for whom we collected the data. It just may be that students who are successful (i.e. high grades and/or achievement of the certificate) using Riff would have been successful anyway because they have prior experience using online learning platforms or they are high-performing students anyway regardless of the learning environment or tools.

Further experiments can be designed as future work to enhance our understanding of the effectiveness of the Riff Platform, and Riff Video in particular. For instance, one can run a controlled trial experiment using Riff Diagnostic in an A/B test setting to see the net effect of the Meeting Mediator nudging tool. An experiment as depicted in Table 5 where, for instance, 4-person groups would work together in a non-learning environment to complete 2 rubric-assessed tasks with treatments would help us identify the net effect of the nudging tool under consideration.

*Table 5. Meeting mediator On or Off during group task*

| Group Type | Task A | Task B |
| --- | --- | --- |
| 1 | On | On |
| 2 | On | Off |
| 3 | Off | On |
| 4 | Off | Off |

In such an experiment, subjects would be compensated for their participation with no specific consequence of performing well or poorly on the tasks (e.g. loss of funds, course failure, etc).

In the case of learning environments, one must take care when setting up control structures such that all students are given equal opportunities for learning. For example, such experiments could attempt to pinpoint relative contributions of different elements of the user-facing application or test different forms and tools of nudging, as selected by students in their personalizations or customizations of the experience. In no case can there be an experimental design in which one group of participants is given a treatment *with the expectation of diminished performance*.

Even with the limitations noted, the experiments we report in our study are a solid first step towards showing that online learning platforms with the right user-facing components that provide relevant and prompt feedback to participants are indeed likely to be effective in enhancing the learning experience. This is an important finding especially in the wake of rapidly growing demand for online learning platforms and tools due to the COVID-19 pandemic. We believe tools like Riff Platform and the Meeting Mediator will have an increasingly important presence in this arena as the educational as well as professional world rely more heavily on these types of platforms.

## 5. Appendix A — Exit Survey Questions and Key Results

**A1. Exit Survey Questions**

1. How well did you understand the lessons at the start of the course?
2. How well did you understand the subject matter at the midpoint of the course?
3. How well did you understand the subject matter at the end of the course?
4. How satisfied were you with what you learned throughout the course?
5. How well did the instructional materials convey course expectations?
6. How well did course assignments test your understanding of course materials?
7. How effectively did the activities and assignments used in this course aid your learning?
8. Would you recommend this course to other students?
9. How well did course staff support your learning throughout the course?
10. How well were your technical questions addressed?
11. How well were your questions about the coding assignments addressed?
12. How well were your questions about the business assignments addressed?
13. How effectively did course staff direct you toward resources?
14. How well did the course staff support you overall?
15. How useful did you find the real-time feedback during video calls?
16. How clear was the real-time feedback during video calls?
17. Thinking about the video calls for your Peer Learning Group and your Capstone team, to what degree do you agree with the following statements?
    - The feedback caused me to participate more.
    - The feedback caused me to give other team members more of an opportunity to speak.
    - The feedback caused me to participate less.
    - The feedback caused the other team members to talk more.
    - The feedback caused the other team members to give other team members more opportunities to speak.
    - The feedback caused the other team members to talk less.
18. On average, approximately what percentage of time were you speaking during meetings?
19. How would you describe how the conversations played out?
20. After using the video tool and seeing the real-time feedback, to what degree do you agree with the following statements?
    - I found this to be a useful tool.
    - I will try to talk more in similar activities in the future.
    - I will try to make sure other people have an equal opportunity to talk.
    - I will try to talk less in similar activities in the future.
    - I would like to use this tool in the future (for learning or work).
21. Do you have any suggestions for how we can improve the real-time feedback?
22. How useful was the Riff Stats feedback?
23. How easy was it to interpret the Riff Stats feedback?
24. How did Riff Stats feedback change your behavior?
25. How useful was the Riff messaging feature (the channels where you connected with other team members)?
26. Considering all the aspects of the Riff Platform (video calling, messaging, real-time feedback, and post-meeting feedback) to what degree do you agree with these statements?

- The Riff Platform helped me connect with my team members.
            - The Riff Platform helped me connect with other people in the course.
            - The Riff Platform helped me connect with course staff.
            - The Riff Platform enabled my team to work effectively together.
            - The Riff Platform enabled me to be more successful in the course.
    27. The Riff Platform was easy to use.
    28. The Riff Platform did not present technical challenges.
    29. The course environment was easy to use.
    30. The course environment did not present technical challenges.
    31. The coding environment was easy to use.
    32. The coding environment did not present technical challenges.
    33. I found it easy to navigate between the course environment, the coding environment, and the Riff Platform.
    34. I was able to use all the tools in the course without any difficulty.

**A2. Key Results**

**General**
73% of course close survey respondents would recommend the course to a friend
89% reported being Satisfied or Very Satisfied with what they learned in the course
63% felt that activities and assignments aided their learning
92% felt well or very well supported by course staff and mentors
NPS = 62

**Collaborative Exercises**
85% of survey respondents reported that the exercises helped them connect with peers
65% of survey respondents reported that the exercises aided their business understanding

**Acknowledgements**

This material is based upon work supported by the National Science Foundation under Grant No. 1843391, SBIR Phase I: Positive Effects of Feedback and Intervention for Engagement in Online Learning.